\documentclass[aps,prl,preprint,groupedaddress]{revtex4-1}

\usepackage{graphicx}
\usepackage{subfigure}
\usepackage{color}
\usepackage{amsmath}
\usepackage[linktocpage,colorlinks=true,linkcolor=blue,citecolor=blue,breaklinks=true]{hyperref}
\usepackage{verbatim}
\usepackage{breakcites}

\newcommand{\ju}[1]{\textcolor{red}{{\it#1}}}

\usepackage{ulem}
\newcommand{\fig}[1]{\textbf{Fig.~\ref{#1}}}

\renewcommand{\vec}[1]{\boldsymbol{#1}}
\newcommand{\tens}[1]{\boldsymbol{#1}}
\newcommand{\bnabla}{\vec{\nabla}}
\begin{document}

\preprint{}

\title{Defect-mediated morphologies in growing cell colonies}

\author{Amin Doostmohammadi}
\affiliation{The Rudolf Peierls Centre for Theoretical Physics, 1 Keble Road, Oxford, OX1 3NP, UK}

\author{Sumesh P. Thampi}
\affiliation{The Rudolf Peierls Centre for Theoretical Physics, 1 Keble Road, Oxford, OX1 3NP, UK}

\author{Julia M. Yeomans}
\affiliation{The Rudolf Peierls Centre for Theoretical Physics, 1 Keble Road, Oxford, OX1 3NP, UK}
\email[Correspondence: ]{j.yeomans1@physics.ox.ac.uk}

\date{\today}

\begin{abstract}
Morphological trends in growing colonies of living cells are at the core of physiological and evolutionary processes. Using active gel equations, which include cell division, we show that shape changes during the growth can be regulated by the dynamics of topological defects in the orientation of cells. The friction between the dividing cells and underlying substrate drives anisotropic colony shapes toward more isotropic morphologies, by mediating the number density and velocity of topological defects. We show that the defects interact with the interface at a specific interaction range, set by the vorticity length scale of flows within the colony, and that the cells predominantly reorient parallel to the interface due to division-induced active stresses. 
\end{abstract}

\pacs{}

\maketitle
Growth dynamics is of considerable importance in biological processes, from biofilm formation \cite{Hall2004} to morphogenesis \cite{Haas2006}, tissue spreading \cite{Ghosh2007}, and tumor invasion \cite{Wolgemuth2011}. 
A prominent feature during the growth of these systems is the emergence of coordinated motion of constituent cells, which may be affected by several mechanisms such as biological signalling between the cells \cite{Haas2006}, chemical cues \cite{Tse2012,Ramin2015}, and mechanical stimuli \cite{TrepatRev}.
Recent experimental studies of bacterial colonies and cellular assemblies show growing evidence of an important role of mechanical factors in regulating growth and in determining collective migration  \cite{Cicuta2014,Rosalind2015a,Rosalind2015b,Trepat2009,Tambe2011,Vedula2012,Doxzen2013,Lene2013,Lene2015}.
In particular, the emergence of the collective motion of cells is often connected to the generation of active stresses by molecular motors and actin polymerisation dynamics and by cell division \cite{Joanny2010,Rossen2014,oursSM2015}. 
Within this context cellular assemblies and bacterial colonies can be modelled as active gels, and the equations of active nematic liquid crystals have been shown to reproduce several experimental observations such as the collective migration of cells \cite{Julicher2008,Julicher2009,Prost2015} and the flow fields of dividing cells \cite{oursSM2015}. 

\begin{figure}[tdp]
\includegraphics[trim = 0 0 0 0, clip, width=0.5\linewidth]{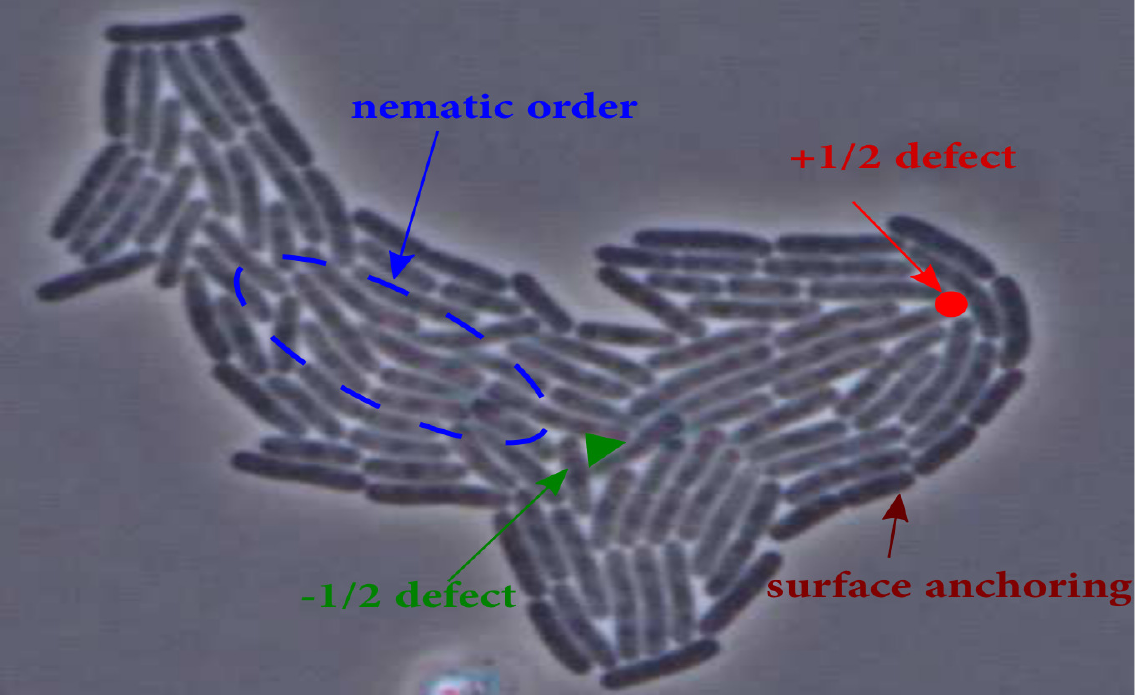}
\caption{A growing {\it E. coli} colony shows nematic ordering, $\pm {1}/{2}$ topological defects and tangential alignment to the interface  
(Picture courtesy of Lin Chao, to whom all rights are reserved).
}
\label{fig:exp}
\end{figure}

The relevance of nematic models is highlighted in \fig{fig:exp}, which shows a snapshot of a dividing {\it E. coli} colony. The orientation field of the rod-shaped {\it E. coli} shows clear local nematic order, corresponding to alignment of the bacteria. 
{\it Topological defects} can also be identified. Such defects cannot be removed by a local realignment of the orientation, and at a defect core the ordering is destroyed.  The strength of a defect is measured as the change in the nematic orientation following a closed curve around the defect core \cite{DeGennesBook}. Therefore $\pm{1}/{2}$ defects, identified in \fig{fig:exp}, correspond to $\pm\pi$ rotations of the bacterial orientation around the defect. Note also the preferential alignment of the cells tangential to the surface. The emergence of nematic order and topological defects have been reported for bacterial colonies \cite{Poon2015}, cultures of fibroblast \cite{Wasoff1976,Silberzan2014}, in living amoeboid cells \cite{Gruler1999}, and more recently in 
Madin-Darby Canine Kidney (MDCK) cells \cite{Benoit}.
~However, to the best of our knowledge, the role of topological defects in growth dynamics and their connections to the morphological responses of cell cultures have not yet been explored.
%

Here, we show how the dynamics of topological defects contribute to shape changes in growing colonies of dividing cells. Building on the active gel description of cellular layers \cite{Prost2015,oursSM2015}, we show that the progression of the interface of a growing colony and its morphological properties are correlated with the generation of defects and their associated dynamics. 
In addition, we relate friction between cells and the underlying substrate to conformational changes based on the increase in the number of defects and reduction of their velocities with increasing friction.
~\\
To represent the dynamics of a growing colony, we use a continuum description of cells as an active gel growing in an  isotropic liquid \cite{Kruse2005,Joanny2007,Volfson2008,MatthewPRL,oursSM2015}. The fields that describe the system are the total density $\rho$, the concentration of cells $\varphi$ which is 1 within the colony and 0 outside, the velocity $\vec{u}$, and the nematic order parameter $\mathbf{Q} = 2q (\mathbf{nn} - \mathbf{I}/2)$, where $ \mathbf{n}$ is the director and $q$ the magnitude of the nematic order.


 The nematic tensor is evolved according to the Beris-Edwards equation \cite{BerisBook} 
\begin{align}
\left(\partial_t + \vec{u}\cdot\bnabla\right) \tens{Q} - \tens{S} &= \Gamma_{Q} \tens{H},
\label{eqn:lc}
\end{align}
where $\tens{S}=\lambda\tens{E}-(\tens{\omega}\cdot\tens{Q}-\tens{Q}\cdot\tens{\omega})$
is a generalised advection term, characterising the response of the nematic tensor to velocity gradients. Here, $\tens{E}=(\bnabla\vec{u}+\bnabla\vec{u}^{T})/2$ is the strain rate tensor, $\tens{\omega} = (\bnabla\vec{u}^{T}-\bnabla\vec{u})/2$ the vorticity tensor, and $\lambda$ is the alignment parameter representing the collective response of cells to velocity gradients. 
$\Gamma_{Q}$ is a rotational diffusivity 
and the molecular field $\tens{H} = -\frac{\delta \mathcal{F}}{\delta \tens{Q}} + \frac{\tens{I}}{3} {\rm Tr} \left(\frac{\delta \mathcal{F}}{\delta \tens{Q}}\right)$, models the relaxation of the orientational order to minimise a free energy $\mathcal{F}_{LC} =  \frac{1}{2}A\left(q_{n}^{2}\varphi-\frac{1}{2}\text{tr}(\tens{Q}^{2})\right)^{2}+\frac{1}{2}K(\bnabla\tens{Q})^{2}$,
where $K$ and $A$ are material constants and the coupling ensures that nematic order with $q=q_{n}$ is favoured for $\phi=1$ and vanishes in the isotropic phase ($\phi=0$). 

%
\begin{figure*}[htdp]
\includegraphics[trim = 0 45 0 0, clip, width=1.0\linewidth]{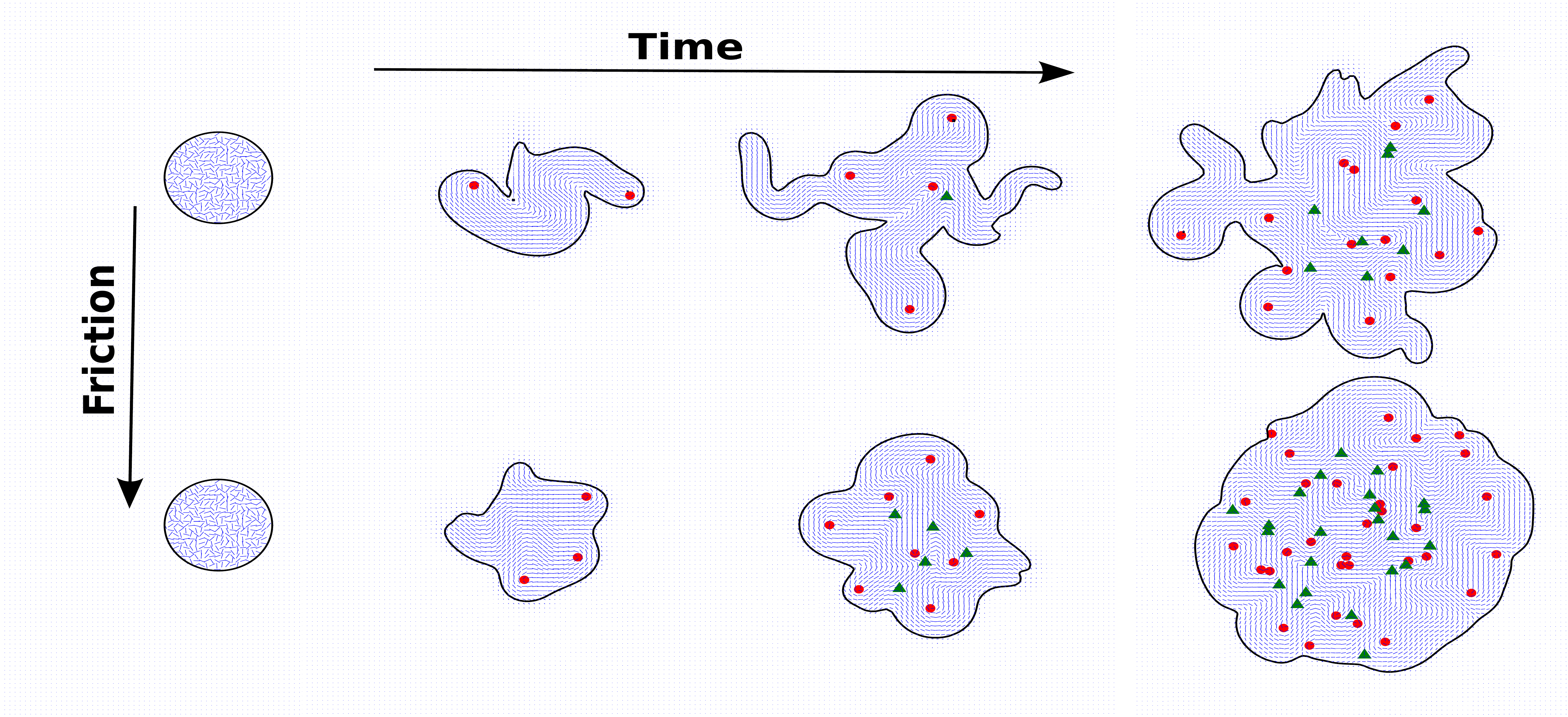}
\caption{The spatio-temporal evolution of a growing colony is affected by the dynamics of topological defects and by friction with the substrate. We show the nematic orientation field, corresponding to local cell directions, (blue solid lines) and $+{1}/{2},-{1}/{2}$ defects marked by red circles and green triangles, respectively.
}
\label{fig:shape}
\end{figure*}
%
To account for the dynamics of the concentration of cells, 
the Cahn-Hilliard equation \cite{Cahn1958} is considered with a growth term to account for cell proliferation
\begin{align}
\partial_{t}\varphi+\bnabla\cdot(\vec{u}\varphi)=\Gamma_{\varphi}\bnabla^{2}\mu+\alpha\varphi,
\label{eqn:conc}
\end{align}
where $\Gamma_{\phi}$ is the mobility, $\alpha$ is the division rate, $\mu=\frac{\delta\mathcal{F}}{\delta\varphi} -\bnabla\varphi\cdot \frac{\delta \mathcal{F}}{\delta \bnabla \varphi}$ is the chemical potential and the free energy functional $\mathcal{F}=\mathcal{F}_{LC}+\mathcal{F}_{GL}$ includes additional contributions from Ginzburg-Landau free energy $\mathcal{F}_{GL} =  \frac{A_{\varphi}}{2}\varphi^{2}(1-\varphi)^{2}+\frac{K_{\varphi}}{2}(\bnabla \varphi)^2$
to allow for phase ordering and surface tension between isotropic and nematic fluids. 
The total density $\rho$ satisfies the continuity equation and the velocity $\bf {u}$ evolves according to 
 \begin{align}
 \rho (\partial_t + \vec{u}\cdot\bnabla)\vec{u} &= \bnabla\cdot\tens{\Pi}-f\vec{u},
\label{eqn:ns} 
 \end{align}
where $\boldsymbol{\Pi}$ denotes the stress tensor and $f=f_{0}\varphi$ with $f_{0}$ the friction coefficient between the cells and the underlying substrate. 
The stress contributions comprise the viscous stress $\tens{\Pi}^{viscous} = 2 \eta\tens{E}$, where $\eta$ is the viscosity,
the elastic stresses $\tens{\Pi}^{elastic}=-P\tens{I} -2\lambda q_{}\tens{H}+ \tens{Q}\cdot\tens{H} - \tens{H}\cdot\tens{Q}-\bnabla \tens{Q} \frac{\delta \mathcal{F}}{\delta \bnabla \tens{Q}}$,
where $P=p-\frac{K}{2}(\bnabla\tens{Q})^2$ is the modified pressure and capillary stresses $\tens{\Pi}^{cap}=(\mathcal{F}-\mu\varphi)\tens{I}-\bnabla \varphi \frac{\delta \mathcal{F}}{\delta \bnabla \varphi}$.

A local increase in the concentration of cells,  driven by the $\alpha\varphi$ term in Eq.~(\ref{eqn:conc}), generates dipole-like flow fields. These reproduce the experimentally measured flow fields of dividing MDCK cells and can be regarded as a source of active stress generation in cell monolayers \cite{oursSM2015}. In the simulations cell division events are introduced randomly across the cell domain as a local increase of $\alpha\varphi$ over a short time of $t_{0}$ time steps, in circles with radius of $r_{0}$ grid points. 
For more details of the method and comparison with experiments see \cite{MatthewPRL,oursSM2015}. 

Eqs.~(\ref{eqn:lc}--\ref{eqn:ns}) were solved numerically using a hybrid lattice Boltzmann method \cite{Denniston2004, Davide2007,Suzanne2011,ourpta2014}. The simulation parameters are $\rho=1$, $\Gamma_{Q}=0.025$, $\Gamma_{\phi}$=0.1, $\kappa=0.05$, $K=0.01$, $\alpha=0.0001$, $\lambda=0.3$, $q_{n}=1$, $A=0.8$, $A_{\phi}=0.6$, $\eta=2/3$, $t_{0}=10$, and $r_{0}=3$, in lattice units. Simulations were performed in a domain of size $800 \times 800$ with an initial circular colony of radius $R_{0}=10$.
%
\begin{figure}[tdp]
\includegraphics[trim = 0 0 0 21, clip, width=0.5\linewidth]{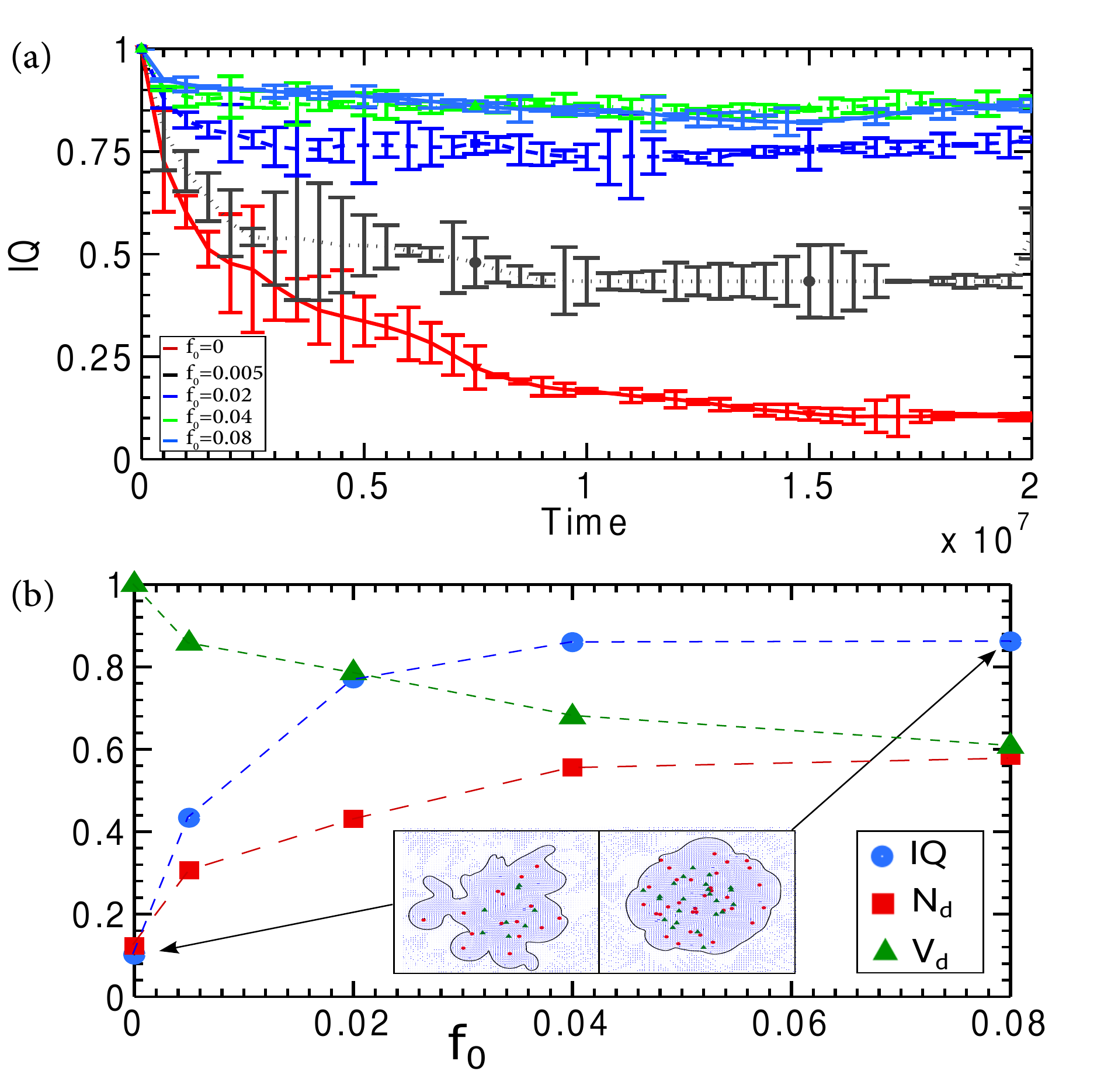}
\caption{Friction controls shape changes in a growing colony by affecting the dynamics of defects. (a) The conformational changes over time (characterised by IQ) as a function of friction. 
(b) Dependence on friction of IQ, the number density of defects N$_{\text{d}}$, and the average defect velocity V$_{\text{d}}$ at long times.}
\label{fig:iq}
\end{figure}
%

An initially circular assembly of dividing cells, with initial random orientations, grows in space as the cells begin to divide. The time evolution of the colony is shown in \fig{fig:shape} for low and high friction. At low friction the colony's shape is characterised by a number of finger-like protrusions reaching into the isotropic liquid. At higher friction the fingering is greatly reduced and the shape of the colony remains close to circular. 

A quantitative measure of the shape change can be obtained by calculating the isoperimetric quotient of the growing colony IQ$=4\pi\mathcal{A}/\mathcal{P}^{2}$, where $\mathcal{A}$ is the surface area and $\mathcal{P}$ is the perimeter of the cell assembly. For a perfect circular geometry IQ$=1$, and it decreases as the shape deviates from a circle. 
At zero friction, the IQ drops rapidly as the initial circular geometry is perturbed by defect-induced protrusions [Fig.~\ref{fig:iq}(a), {\it red data}] before flattening to an asymptotic value for longer times. The higher IQ measured for larger friction indicates the tendency towards forming a more circular morphology [Fig.~\ref{fig:iq}(a), {\it cyan line}].

The positions of $\pm1/2$ topological defects are also shown in \fig{fig:shape}. 
The defects can be identified by calculating the diffusive charge density \cite{MatthewPRL}, $s=\frac{1}{2\pi}\left(\frac{\partial Q_{xx}}{\partial x}\frac{\partial Q_{xy}}{\partial y}-\frac{\partial Q_{xx}}{\partial y}\frac{\partial Q_{xy}}{\partial x}\right)$,
which gives $s=\pm 1/2$ at the position of defect cores.
It is apparent from \fig{fig:shape}, and known from previous work \cite{SumeshPRE2014,oursNM2015}, that the number of defects in an active nematic increases with friction [\fig{fig:iq}(b), {\it red data}]. Moreover the average defect velocity V$_{\text{d}}$ decreases with increasing friction [\fig{fig:iq}(b), {\it green data}].
%
\begin{figure*}[tdp]
\includegraphics[trim = 0 0 0 0, clip, width=1.0\linewidth]{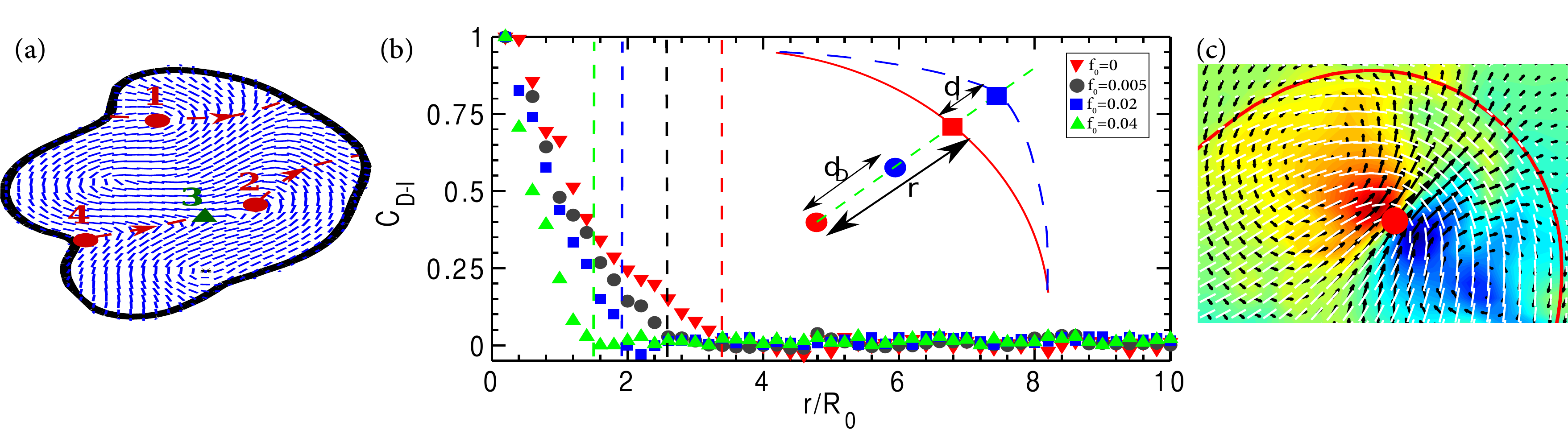}
\caption{The displacement of the interface is correlated with the motion of nearby defects. (a) The typical trajectories of topological defects created at the interface (1,4) or as a pair in the bulk (2--3). The director field is shown by blue solid lines and $+{1}/{2},-{1}/{2}$ defects are illustrated by red circles and green triangles, respectively. (b) The defect-interface cross correlation for different frictions versus distance normalised by the initial radius of the colony $R_{0}$ and (c) the flow field around a defect moving towards the interface. Vertical dashed lines in (b) correspond to the characteristic vorticity length scale at different frictions. 
The inset in (b) shows a schematic of the displacements of a defect (circle) and the interface (line). The position of the nearest point at the interface along the direction of the defect velocity is marked by a square. Red and blue colors correspond to consecutive  positions at times $t$ and $t+1$, respectively. 
The colormap in (c) illustrates the vorticity contours, superimposed by velocity vectors (black arrows) and cell orientations (white solid lines).}
\label{fig:corr}
\end{figure*}

Figure \ref{fig:corr}(a) shows in more detail how defects are created and annihilated. Defects can either be created at the interface (defects 1 and 4) due to surface undulations, leaving a net negative charge on the interface, or appear as a pair in the bulk (defect pair 2--3) to release elastic energy associated with the orientation field. Once created, comet-like $+1/2$ defects move through the colony and either annihilate with the oppositely charged defects in the bulk (defect 4 annihilates with defect 3) or approach the interface (defects 1 and 2). 
As a defect moves towards the interface the dynamic progression of the border adapts to the motion of the approaching defect until it reaches the surface and annihilates with the negative charge density that is distributed along the interface. This suggests that the shape of the colony will be correlated to the motion of the defects and it is apparent from \fig{fig:exp} and \fig{fig:shape} that finger-like protrusions are closely linked to the motion of comet-like (+1/2) topological defects.

When the friction between cells and the underlying substrate is small, topological defects moving towards the border and the consequent protrusions are few in number and energetic,  and the shape of a growing assembly is highly anisotropic [Fig.~\ref{fig:shape}, {\it top row}]. However, as the friction is enhanced, the number of topological defects increases and the average defect velocity is reduced resulting in a more isotropic morphology [Fig.~\ref{fig:shape}, {\it bottom row}]. 

To obtain a more quantitative characterisation of the connection between the movement of topological defects and the progression of the interface, we define the cross-correlation function between defect and interface velocities $C_{D-I}(r)=\langle \vec{v}_{D}(r,t)\cdot \vec{v}_{I}(r,t)\rangle/\langle \vec{v}_{D}(0,t) \cdot \vec{v}_{I}(0,t)\rangle$, where $\vec{v}_{D}=\Delta d_{D}/\Delta t$ denotes the defect velocity and $\vec{v}_{I}=\Delta d_{I}/\Delta t$ is the velocity of the point on the interface at a distance $r$ from the defect  in the direction of $\vec{v}_{D}$ [see Fig.~\ref{fig:corr}(b), {\it inset}]. The measurement of this cross-correlation function demonstrates that the defect and interface displacements are correlated over a given distance, which we term the {\it interaction range} [Fig.~\ref{fig:corr}(b)]. The interaction range decreases as the friction is increased. 

To determine the physical mechanism for the emergence of the interaction range and its dependence on the friction, we consider the flow field and director configuration around a defect approaching the interface [Fig.~\ref{fig:corr}(c)]. As evident from the figure the motion of the defect sets up counter-rotating velocity vortices \cite{Julia2002,ourprl2013,Giomi2013,Giomi2014}. 
We conjecture that the interaction range is controlled by the vorticity length scale in the growing colony. To show that this is indeed the case we calculate the  characteristic vorticity length scale from the vorticity-vorticity correlation function and compare it to the interaction range for different frictions [Fig.~\ref{fig:corr}(b), {\it dashed lines}]. The close agreement between the length scales associated with the vorticity and the defect-interface interaction range shows that the defect motion towards the interface affects the interface deformation through vortex generation around a defect. As the friction is increased the vorticity field is more effectively suppressed by hydrodynamic screening and therefore the characteristic vorticity length is reduced \cite{oursNM2015}, leading to a shorter interaction range between defects and the interface [Fig.~\ref{fig:corr}(b)]. 

\begin{figure}[hdp]
\includegraphics[trim = 0 10 0 0, clip, width=0.5\linewidth]{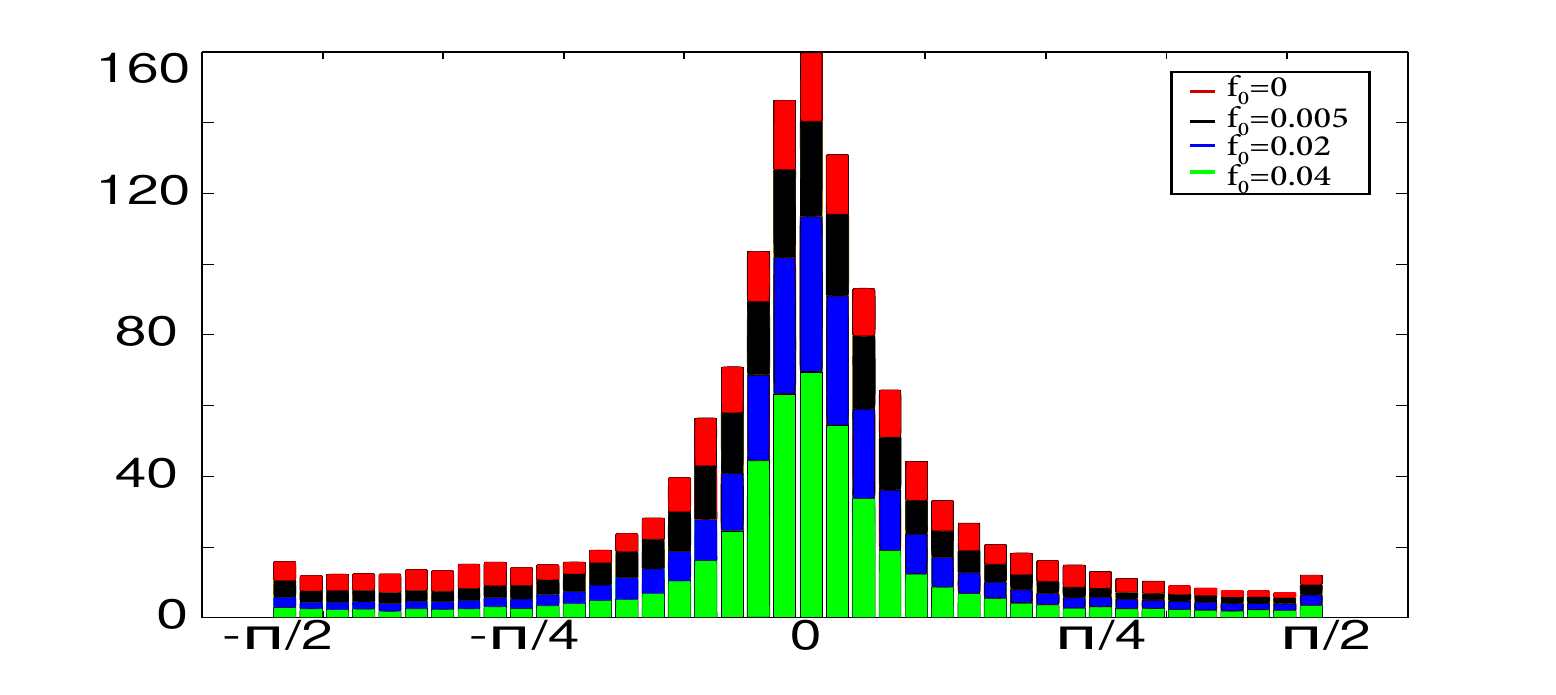}
\caption{Dividing cells align along the free surface: distribution of anchoring angle relative to the surface for different frictions.}
\label{fig:anch}
\end{figure}

An interesting feature observed during the simulated growth of the cell colony is that cells at the border tend to lie parallel to the interface. This is quantified by plotting the distribution of the anchoring angle relative to the interface of the outermost cells [Fig.~\ref{fig:anch}]. Similar behaviour is evident in \fig{fig:exp} and in the experiments reported in \cite{Volfson2008}. In a recent study of 
lyotropic active nematics Blow {\it et al.} \cite{MatthewPRL} showed that the gradients in order and orientation of nematogens along an interface result in the generation of activity-induced anchoring forces which, for extensile active nematics, induce tangential orientations along the interface. Since the cell division generates extensile stresses \cite{oursSM2015}, we expect the same mechanism to be responsible for the anchoring along the border of a growing colony.

It has been shown recently that by increasing the adhesion between the cells and substrate the spreading of cell aggregates is enhanced: groups of cells are formed in the shape of finger-like structures leading to anisotropic shapes \cite{Ravasio2015}. This was interpreted in terms of `leader' cells at the end of the fingers. By contrast previous theoretical predictions have associated the fingering to the curvature dependent motility of cells \cite{Silberzan2010} or undulation instabilities due to cell proliferation in epithelial tissues \cite{Basan2011,Prost2015b}, neglecting the orientational order of cells. Here, we offer an alternative, collective physical mechanism for such morphological changes based on the dynamics of topological defects in a growing colony. 
To test these ideas the friction between cells and their underlying substrate could be used to vary the number of defects in active systems \cite{SumeshPRE2014,oursNM2015,Francesc2015}.  A potential approach would be to consider a fixed amount of Fibronectine (FB) proteins, which control cell binding to the substrate, and explore the spreading dynamics on substrates with varying stiffness to introduce differing hydrodynamic screening \cite{Benoit}. 
Indeed recent experiments do show the role of friction in pattern formation in active matter \cite{Hannezo2015,Francesc2015} and demonstrate important connections between the friction and cell motility \cite{Paluch2015}.   
We have described a new physical mechanism for understanding morphological changes in growing colonies based on the dynamics of topological defects in cell orientations. 
Such defects have been recently observed experimentally \cite{Poon2015,Benoit} and our results suggest directions for further investigations of defect-mediated migration and morphologies in cellular colonies.
\begin{acknowledgments}
We acknowledge funding from the ERC Advanced Grant (MiCE 291234). We thank Matthew Blow, Benoit Ladoux, Wilson Poon, Thuan Beng Saw, Tyler Shendruk, and Ben Simons for helpful discussions.
\end{acknowledgments}
%
\bibliography{refe.bib}

\end{document}